\documentclass{aa}
\def\mdot {\dot M}

\def\ergs {~erg$\,$s$^{-1}$}

\def\cmdue {~cm$^{-2}$}

\def\msole{~M_{\odot}}

\begin{document}

\def\gsim{ \lower .75ex \hbox{$\sim$} \llap{\raise .27ex \hbox{$>$}} } 
\def\lsim{ \lower .75ex\hbox{$\sim$} \llap{\raise .27ex \hbox{$<$}} } 

\input{psfig.tex}

\title{BeppoSAX observation of Cen X-4 in quiescence}

\author{Sergio Campana$^1$, Luigi Stella$^2$, Sandro Mereghetti$^3$, 
Davide Cremonesi$^3$}

\institute{
Osservatorio Astronomico di Brera, Via Bianchi 46, I--23807 
Merate, Italy
\and
Osservatorio Astronomico di Roma, Via Frascati 33,
I-00040 Monteporzio Catone (Roma), Italy
\and 
Istituto di Fisica Cosmica ``G. Occhialini", CNR, Via Bassini 15, 
I--20133 Milano, Italy
}

\date{Submitted ..... / Accepted .....}
\offprints{campana@merate.mi.astro.it}
\maketitle
\markboth{Campana et al.}
         {BeppoSAX observations of Cen X-4 in quiescence}

\begin{abstract}
We report on a 61 ks BeppoSAX observation of the soft X--ray 
transient Cen X-4 during quiescence which allowed to study the source 
X--ray spectrum from $\sim 0.3$ keV to $\sim 8$ keV. A two-component 
spectral model was required, consisting of a black body with temperature 
of $\sim 0.1$ keV and a power law with photon index $\sim 2$. 
These values are compatible with earlier ASCA results indicating that 
Cen X-4 may be stable, within a factor of a few, over a 5 year baseline.
\end{abstract}


\section{Introduction}

Cen X-4 is one of the best studied sources of the Soft X--Ray Transient 
(SXRT) class (for a review see Campana et al. 1998a). X--ray outbursts 
have been detected in 1969 and 1979. 
During the 1979 outburst Cen X-4 reached a peak flux of $\sim 5$~Crab, 
corresponding to $L_X\sim 4\times 10^{37}$\ergs\ for a distance of
$d \sim 1.2$~kpc (Kaluzienski, Holt \& Swank 1980). 
Type I bursts were observed, testifying 
to the presence of an accreting neutron star.

During the 1979 outburst the optical counterpart could be identified, 
as it brightened by $\gsim 6$ magnitudes (Caniza\-res, McClintock \& 
Grindlay 1979).
Extensive spectroscopic and photometric measurements of the optical
counterpart in quiescence (V=18.7 mag) led to the determination of
the orbital period (15.1 hr; Chevalier et al. 1989;
McClintock \& Remillard 1990) and mass
function ($\sim 0.2 \msole$, converting to a neutron star mass between
$0.5-2.1 \msole$). The optical spectrum 
shows the characteristics of a K5-7 main sequence star, contaminated by
lines (e.g. H$\alpha$, H$\beta$ and H$\gamma$) and continuum emission 
probably resulting from an accretion disk (Shahbaz, Naylor \& Charles 1993). 
The latter was estimated to contribute $\sim 80\%$, $\sim 30\%$, $\sim 
25\%$ and $\sim 10\%$ of the quiescent optical flux in the B, V, R and 
I bands, respectively.

ASCA observed twice Cen X-4 in its quiescent state in Feb. 1994 (28 ks) 
and Feb. 1997 (39 ks). 
In the first observation Cen X-4 was detected at a luminosity of $L_X 
\sim 2^{+2}_{-1}\times 10^{32}$\ergs\ (0.5--10 keV, Asai et al. 1996, 1998). 
The X--ray spectrum was well fit by a black body component 
($k\,T_{\rm bb}=0.16^{+0.03}_{-0.02}$ keV) plus an additional power-law 
component with photon index $\Gamma = 1.9\pm0.3$. The 0.5--10 keV flux 
from the two spectral components was comparable. 
The column density was constrained to be $N_H \lsim 2\times 
10^{21}$\cmdue. The equivalent radius of the black body component was 
determined to be $\sim 1.8$ km, substantially smaller than the radius of a 
neutron star. A search for X--ray pulsations gave negative results, 
providing an upper limit to the pulsed fraction of $\sim 50\%$ for periods 
between 8 ms and 8200 s (Asai et al. 1996). 
The second ASCA observation provided similar results: 
$L_X \sim 3^{+3}_{-2}\times 10^{32}$\ergs, $k\,T_{\rm bb}=0.13\pm0.02$ keV,
$\Gamma = 2.5\pm0.5$ and $N_H = (3\pm1) \times 10^{21}$\cmdue\ 
(Asai et al. 1998).

During quiescence Cen X-4 was also observed with the  
Einstein IPC (in 1980, $\sim 440$ d after the 1979 outburst; Petro et
al. 1981), EXOSAT CMA (in 1986; van Paradijs et al. 1987) and 
ROSAT HRI (in 1995; Campana et al. 1997).
Campana et al. (1997) showed that, taking into account the relative 
uncertainties, these measurements are consistent with the same luminosity 
level of the ASCA observations.
Yet, during the ROSAT HRI pointing a factor of $\sim 3$ flux variation 
was observed on a timescale of a few days (Campana et al. 1997).

Here we report on a BeppoSAX observation of Cen X-4 in its quiescent state. 
We describe the spectral and timing analysis
of the BeppoSAX data as well as a reanalysis of the ASCA data in Section 2. 
In Section 3 we discuss our results.

\begin{figure*}[!htb]
\psfig{file=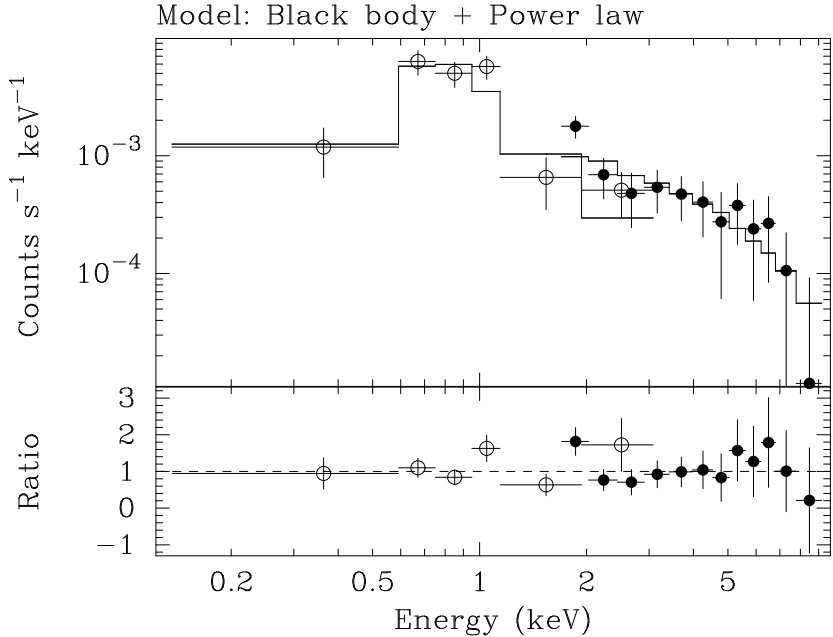,width=8truecm}
\caption[h]{LECS (circles) and MECS (dots) X--ray spectrum of Cen X-4 
in quiescence. 
The spectrum is fit with a black body plus a power law model. 
In the lower panel the ratio between the data and the model 
is presented.}
\end{figure*}

\section{BeppoSAX observation}

A BeppoSAX (Boella et al. 1997a) observation of Cen X-4 took place on 
9--11 Feb. 1999 for a total elapsed time of 135 ks. 
The source was detected only by the Low Energy Concentrator Spectrometer 
(LECS; 0.1--10 keV, Parmar et al. 1997) and the Medium Energy  Concentrator
Spectrometer (MECS; 1.3--10 keV, Boella et al. 1997b).
Due to the South Atlantic Anomaly and Earth occultations 
the net exposure was 61 ks with the MECS and 21 ks with the LECS.
The latter instrument could be operated only during the satellite night-time.
 
\subsection{Spectral analysis}

The LECS and MECS events were extracted within a radius of $4'$ centered 
on the source position. We collected 233 photons from 
the LECS and 632 from the MECS in the full energy range. 
Background subtraction was applied using 
the standard BeppoSAX files. 
The source background subtracted count rates were $(4.7\pm 0.6) \times 
10^{-3}$ (0.1--3.1 keV) and $(2.6\pm 0.4) \times 10^{-3}$ ct s$^{-1}$ 
(1.7--9.0 keV) in the LECS and MECS instruments, respectively. 
The LECS and MECS spectral data were rebinned 
in order to have at least 40 photons per channel. The spectral analysis was 
carried out in the energy range relative to the count rates.
A variable normalization factor was included to
account for the mismatch in the absolute flux calibration of the BeppoSAX
instruments. This factor is usually in the 0.7--1 range and has a value
of $\sim 0.7$ in the best fit discussed below.

\begin{figure*}[!htb]
\psfig{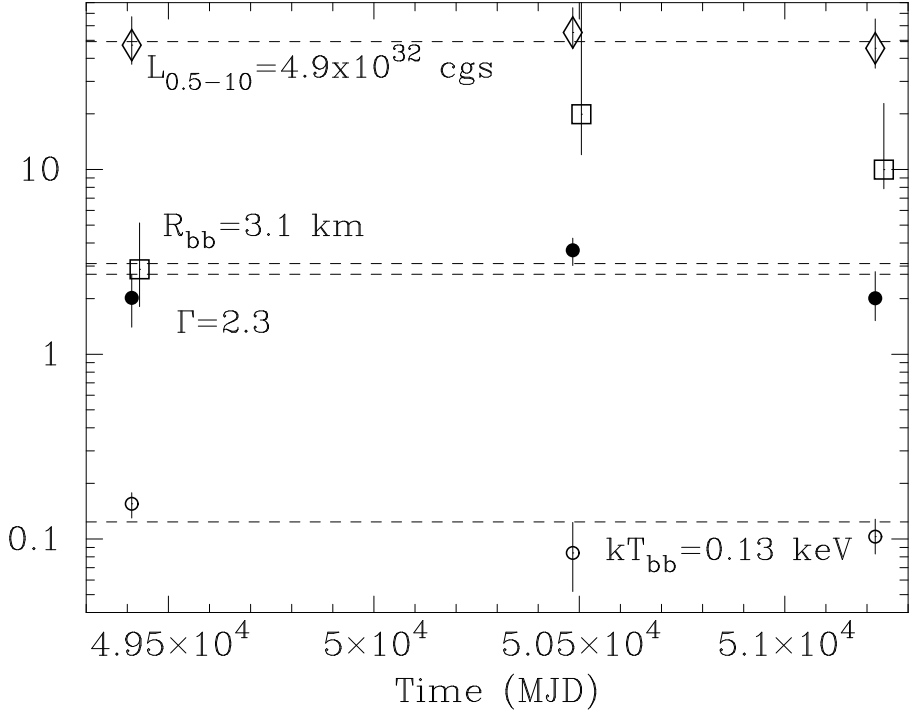}
\caption[h]{Long term behaviour of the overall 0.5--10 keV 
unabsorbed luminosity in units of $10^{31}$\ergs\
(diamonds), of the power law photon index (filled circles), of the 
equivalent black body radius in km (squares, slightly offset in time)
and of the equivalent black body temperature in keV (open circles), 
taken with ASCA and BeppoSAX imaging instruments. Errors are 
at the 68\% confidence level for three parameters of interest ($\Delta 
\chi^2=3.5$).
These uncertainties have been adopted to derive the mean in Table 1.
The errors on luminosity has been taken from Asai et al. 
(1996, 1998). In the case of the BeppoSAX observation we assume a 50\% error.
Luminosities are unabsorbed and in the energy range 0.5--10 keV.
}
\end{figure*}

We fit the spectral data with the XSPEC (version 10.00) package.
All-single component models provided a poor fit to the data ($\chi^2_{\rm red} 
\gsim 2$). Therefore we applied the conventional model for quiescent 
neutron star SXRTs, i.e. a soft black body component plus a hard power-law 
tail (Asai et al. 1996, 1998; Campana et al. 1998a,b). The fit was good 
($\chi^2_{\rm red}=1.1$; cf. Figure 1).
The corresponding black body temperature was $k\,T_{\rm bb}=103^{+53}_{-32}$ 
eV and the power law photon index $\Gamma=2.01^{+0.65}_{-0.68}$ 
(uncertainties are 90\% confidence for a single parameter, $\Delta 
\chi^2=2.71$). 
The equivalent black body radius was $R_{\rm bb}=10.0^{+420}_{-2.2}$ km. 
Note that this radius, though highly uncertain, is consistent with the 
neutron star radius. A $3\,\sigma$ lower limit of $R_{\rm bb}\gsim 6.2$ km
on the emitting region was derived. 
The column density was $N_H=2.6^{+5.2}_{-2.3}\times 10^{21}$\cmdue.
The absorbed X--ray flux was $7.3\times 10^{-13}$ and $2.0\times 
10^{-13}$\ergs\cmdue\ in the 0.1--2 and 2--10 keV energy
bands, respectively. The corresponding unabsorbed flux was $8.5\times 
10^{-12}$ and  $2.0\times 10^{-13}$\ergs\cmdue, respectively,
providing (unabsorbed) luminosities of 
$1.5\times 10^{33}$\ergs\ (0.1--10 keV) and $4.5\times 10^{32}$\ergs\
(0.5--10 keV). 

\subsection{Timing analysis}

The relatively small number of counts did not permit to carry out 
a sensitive search for periodicity and short term variability. 
The MECS and LECS light curves were consistent with a constant flux.
A variation by a factor of $\sim 3$ on a timescale comparable to the 
observation duration (i.e. similar to that observed with the ROSAT HRI, see
Campana et al. 1997) would have been easily detected in the BeppoSAX light curves.

\subsection{Reanalysis of ASCA data}

The large equivalent black body radius derived from BeppoSAX data 
($R_{\rm bb}\gsim 6.2$ km) is larger than that  
reported by Asai et al. (1996, 1998) using ASCA data.
Since these authors do not report the uncertainties on the equivalent 
black body radius, we reanalysed the ASCA data. 
We consider only GIS data since Cen X-4 falls across two CCDs in the SIS.   
The best fit values for the black body plus power law model
are given in Table 1. In the first observation the best fit value of 
the column density is consistent with zero and the other spectral parameters 
are consistently different from the values derived from the second ASCA 
and BeppoSAX observations.
In the second ASCA observation, the count rates of all the ASCA instruments 
are systematically lower than those in the first observation, despite the 
similar source position on the detectors, with a lack of photons at soft energies.
For this observation, we derive an (unabsorbed) X--ray luminosity which is 
a factor of $\sim 3$ higher than the value quoted by Asai et al. (1998).

The large uncertainties in the parameters are signs of poor statistics. 
Due to these uncertainties, we also repeated the 
spectral analysis keeping the column density value fixed to the BeppoSAX value 
($2.6\times 10^{21}$\cmdue). These fit provide only a marginally worse description 
of the data and result in a much more homogeneous spectral parameter values across
the different observations.

%

\begin{table*}
\caption{Summary of Cen X-4 spectral fit. For the ASCA data we consider only the GIS 
detectors. Free column densities are considered in the upper part of the table. 
Column densities fixed to the best fit BeppoSAX value are reported in the lower part.
Uncertainties are for one parameter of interest and at the 90\% confidence level 
($\Delta \chi^2=2.71$)}
\begin{tabular}{ccccccc}
Date      & Column density & Photon index         & Temperature          & Radius            & Luminosity             & Red. $\chi^2$ \\ 
(dd/mm/yy)&($10^{21}$ cm$^{-2}$)&                 & (keV)                & (km)              & ($10^{32}$\ergs)       &               \\
\hline
27/02/94 &$0.2^{+5.3}_{-0.2}$&$1.7^{+0.9}_{-0.6}$&$0.19^{+0.03}_{-0.07}$&$\ 1.1^{+8.0}_{-0.3}$&   $2.6$              & 1.0 \\
04/02/97 &$3.5^{+4.8}_{-3.5}$&$3.7^{+0.8}_{-0.9}$&$0.08^{+0.14}_{-0.02}$&$29.4^{+287}_{-29.4}$&   $8.6$              & 1.1 \\
09/02/99 &$2.6^{+5.2}_{-2.3}$&$2.0^{+0.7}_{-0.7}$&$0.10^{+0.05}_{-0.03}$&$10.0^{+420}_{-2.2}$ &   $4.5$              & 1.1 \\
\hline
\hline
27/02/94 &2.6 (fixed)        &$2.0^{+0.6}_{-0.5}$&$0.15^{+0.02}_{-0.02}$&$\ 2.9^{+1.9}_{-1.0}$&   $4.7$              & 1.0 \\
04/02/97 &2.6 (fixed)        &$3.6^{+0.5}_{-0.5}$&$0.08^{+0.03}_{-0.03}$&$18.9^{+145}_{-6.3}$ &   $5.5$              & 1.1 \\
09/02/99 &2.6 (fixed)        &$2.0^{+0.7}_{-0.5}$&$0.10^{+0.02}_{-0.02}$&$10.0^{+10.8}_{-1.8}$&   $4.5$              & 1.1 \\
\hline
MEAN$^\ddag$     &  --               &$2.7\pm0.7$ (1.4\%)&$0.12\pm0.03$ (2.7\%)& $3.1^{+3.7}_{-3.1}$ (74\%)&$4.9\pm1.9$ (84\%) &  \\ 
\hline
\end{tabular}

\noindent $^\ddag$ Mean of the three observations in the case of 
fixed column density. In parenthesis are also reported the null hypothesis 
probabilities relative to the fit with a constant function for one degree of 
freedom. The mean has been computed using larger uncertainties than the ones 
reported in the Table (cf. Fig. 2).

\end{table*}

\section{Discussion}

There is growing evidence that the quiescent spectrum of SXRTs is 
made of two components: a soft component, usually modeled with a 
$k\,T_{\rm bb}\sim 0.1-0.3$ keV black body, and a hard power law 
component with photon index $\sim 1.5-2$ (Campana et al. 1998a,b;
Asai et al. 1996, 1998; Guainazzi et al. 1999). 
The soft component is usually attributed to the radiative cooling 
of the neutron star warm interior heated up during the accretion episodes 
giving rise to the outbursts (Campana et al. 1998a; Brown, Bildsten \& 
Rutledge 1998). Modeling this soft component with a black body, the derived 
emitting area is usually smaller than the neutron star surface, with $R_{\rm bb}\sim 
1-2$ km (Verbunt et al. 1994; Campana et al. 1998a,b; Asai et al. 1996, 1998).
Neutron star atmosphere models have been used to fit the available 
data. These models provide a good fit to the soft component 
and substantially a larger effective temperatures and radii with 
respect to a pure black body model (Rutledge et al. 1999).
The radii inferred in this way are consistent with emission from the 
entire neutron star surface.
The hard component has been interpreted as due to the interaction of 
a relativistic radio pulsar wind with matter outflowing from the companion (shock 
emission), since when the SXRTs set down to quiescence the neutron star may resume 
its activity as a radio pulsar (Stella et al. 1994; Campana et al. 1998a).
Heating of the polar caps by high energy particles responsible 
for the radio pulsar emission may also contribute to the soft component
(e.g. Becker \& Tr\"umper 1999; Campana \& Stella 2000).
Recently, a UV spectrum of Cen X-4 has been obtained with HST/STIS (McClintock 
\& Remillard 1999). The main result is that in the $\nu$ versus 
$\nu\,F_{\nu}$ plot the unabsorbed flux decreases by only a factor of 2 from 
X--rays to optical (subtracted from the contribution of the companion star). 
Such a nearly $\nu\,F_{\nu}$ flat spectrum is clearly reminiscent {\rm of the 
extended power law spectra that are characteristics} of shock emission. 
 
An alternative explanation attributes the soft component to matter 
accretion onto the neutron star surface in the propeller regime. 
In this case a small fraction of the mass inflow 
rate would penetrate the magnetosphere at high neutron star latitudes. 
The hard component instead would be produced in an advection-dominated 
accretion flow (ADAF, Zhang et al. 1998; Menou et al. 1999).
At the moment this is just a suggestion, lacking self-consistent ADAF 
models able to fit the multi-wave\-lenght spectra of SXRTs in quiescence.

In principle, an accurate monitoring of the variability on different
time scales could help in discriminating between these two
possibilities. In fact, while in general accreting binaries show 
flux variations, in the case of cooling plus shock emission model we 
expect a more steady luminosity.

The BeppoSAX observation presented here allowed us to further study the 
quiescent state of Cen X-4. A comparison in the 0.5--10 keV energy range
(i.e. including the hard spectral component) can be carried out only
with the two ASCA observations. 
In the upper part of Table 1 we report the fit with free column 
densities. We obtain large variations in the spectral parameters.
On the other hand if we fix the column density to the BeppoSAX value, 
we obtain similarly good fits and spectral parameters in much more
agreement (lower part of Table 1).
We are unable to decide which of the two options is correct based on 
pure statistics. Given the low number of counts, we should prefer
the fit with constant column density, since the goodness of the fit is
not particularly affected.
We remark however that a count rate percentual variation in the soft 
part of the spectrum by $\sim 30\%$ is observed in the two ASCA observations. 

Together with the nearly constant behaviour of the source flux (within a 
factor of a few in the 0.5--10 keV energy band) on timescales of years 
variations by a factor $\gsim 3$ in a few days in the 
soft (0.1--2.4 keV) energy band (Campana et al. 1997) and variations 
by up to $\sim 1$ mag in the optical (Chevalier et al. 1989; Cowley et al. 
1988) have been observed. The interpretation of these variations is not 
simple based on current models.
In principle, mechanisms based on accretion can more readily explain 
variations in the X--ray luminosity (especially in the ADAF  
regime where the efficiency is proportional to $\mdot^2$).
Variations in the cooling component are not expected whereas the efficiency 
of the shock emission component should be almost constant at least in the soft 
X--ray band. In this scenario more promising mechanisms should involve a 
variation in the enshrouding geometry in proximity of the source and 
therefore of the absorbing column density. 
 
A different interpretation relies on the coronal activity of the companion 
star.
This mechanism provides too low a quiescent luminosity to power SXRTs in 
quiescence, achieving the saturation level of $10^{29}-10^{30}$\ergs\ for 
main sequence stars (Eracleous et al. 1991), unless subgiant companions 
are present ($10^{31}-10^{32}$\ergs\ for RS CVn systems; Campana \& Stella 
2000; Bildsten \& Rutledge 2000).
Even if the basal coronal emission is much lower than the quiescent 
luminosity observed in SXRTs in quiescence, large flare events may reach 
peak 0.1--2.4 keV luminosities in excess of a few $10^{32}$\ergs\ as in 
the RS CVn stars UX Ari ($\sim 2\times 10^{32}$\ergs; G\"udel et al. 1999) 
or Algol ($\sim 2\times 10^{32}$\ergs; Ottmann \& Schmitt 1996). 
Given the short orbital period of Cen X-4 (15.1 hr) compared with RS CVn 
binaries (1--20 d) even more energetic flare might be expected.

\begin{acknowledgements}
We acknowledge useful comments by an anonymous referee and by G. Tagliaferri 
who help improving the discussion.
This research has made use of SAXDAS linearized and cleaned event
files (Rev.2.0) produced at the BeppoSAX Science Data Center.
This research has made use of data obtained through the High Energy
Astrophysics Science Archive Research Center (HEASARC), provided by
NASA's Goddard Space Flight Center.
This work was partially supported through ASI grants.
\end{acknowledgements}


\begin{thebibliography}{99}

\bibitem{ } Asai K. et al., 1996, PASJ 48 257

\bibitem{ } Asai K. et al., 1998, PASJ 50 611

\bibitem{ } Becker W., Tr\"umper J.,  1999, A\&A 341 803

\bibitem{ } Bildsten L., Rutledge R.E., 2000, ApJ submitted (astro-ph/9912304)

\bibitem{ } Boella G. et al., 1997a, A\&AS 122 299

\bibitem{ } Boella G. et al., 1997b, A\&AS 122 327

\bibitem{ } Brown E.F., Bildsten L., Rutledge R.E., 1998, ApJ 504 L95

\bibitem{ } Campana S., Stella L., 2000, ApJ submitted

\bibitem{ } Campana S. et al., 1997, A\&A 324 941

\bibitem{ } Campana S. et al., 1998a, A\&A Rev. 8 279

\bibitem{ } Campana S. et al., 1998b, ApJ 499 L65

\bibitem{ } Canizares C.R., McClintock J.E.,  
Grindlay J.E., 1979, ApJ 234 556

\bibitem{ } Chevalier C., Ilovaisky S.A., van Paradijs J., 
Pedersen H., van der Klis M., 1989, A\&A 210 114

\bibitem{ } Cowley A.P. et al., 1988, AJ 95 1231

\bibitem{ } Eracleous M. et al., 1991, ApJ 382 290

\bibitem{ } Guainazzi M. et al., 1999, A\&A 349 819

\bibitem{ } G\"udel M. et al., 1999, ApJ 511 405

\bibitem{ } Kaluzienski L.J., Holt S.S., Swank J.H., 1980, ApJ 241 779

\bibitem{ } K\"urster M., Schmitt J.H.M.M., 1996, A\&A 311 211

\bibitem{ } McClintock J.E., Remillard R.A., 1990, ApJ 350 386 

\bibitem{ } McClintock J.E., Remillard R.A., 2000, ApJ 531 956

\bibitem{ } Menou K. et al., 1999, ApJ 520 276

\bibitem{ } Ottmann R., Schmitt J.H.M.M., 1996, A\&A 307 813

\bibitem{ } Parmar A.N. et al., 1997, A\&AS 122 309

\bibitem{ } Petro L.D., Bradt H.V., Kelley R.L., Horne K., 
Gomer R., 1981, ApJ 251 L7

\bibitem{ } Rutledge R.E., Bildsten L., Brown E.F., Pavlov G.G., 
Zavlin V.E., 1999, ApJ 514 945

\bibitem{ } Shahbaz T., Naylor T., Charles P.A., 1993, MNRAS 265 655

\bibitem{ } Stella L., Campana S., Colpi M., Mereghetti S., Tavani M., 
1994, ApJ 423 L47

\bibitem{ } van Paradijs J., Verbunt F., Shafer R.A., 
Arnoud K.A., 1987, A\&A 182 47 

\bibitem{ } Verbunt F., Johnston H., Hasinger G.,  Belloni T., Bunk W., 
1994, A\&A 285 903

\bibitem{ }
Zhang S.N., Yu W., Zhang W.W., 1998, ApJ 494 L71

\end{thebibliography}
\end{document}